# Thermal effects on lattice strain in ε-Fe under pressure

Xianwei Sha and R. E. Cohen

Carnegie Institution of Washington, 5251 Broad Branch Road, NW, Washington DC 20015, USA

We compute the c/a lattice strain versus temperature for nonmagnetic hcp iron at high pressures using both first-principles linear response quasiharmonic calculations based on the full potential linear-muffin-tin-orbital (LMTO) method and the particle-in-cell (PIC) model for the vibrational partition function using a tight-binding total-energy method. The tight-binding model shows excellent agreement with the all-electron LMTO method. When hcp structure is stable, the calculated geometric mean frequency and Helmholtz free energy of ε-Fe from PIC and linear response lattice dynamics agree very well, as does the axial ratio as a function of temperature and pressure. On-site anharmonicity proves to be small up to the melting temperature, and PIC gives a good estimate of its sign and magnitude. At low pressures, ε-Fe becomes dynamically unstable at large c/a ratios, and the PIC model might fail where the structure approaches lattice instability. The PIC approximation describes well the vibrational behavior away from the instability, and thus is a reasonable approach to compute high temperature properties of materials. Our results show significant differences from earlier PIC studies, which gave much larger axial ratio increases with increasing temperature, or reported large differences between PIC and lattice dynamics results.

PACS number(s): 65.40.-b, 74.25.Kc, 71.20.Be, 52.65.Rr

# I. Introduction

Zero-temperature properties of materials can be obtained in a straightforward way using first-principles electronic structure methods, but thermal properties are much more difficult to obtain from first principles. Five approaches have been used: application of the phenomenological Debye model, quasi-harmonic lattice dynamics,[3-8] molecular dynamics,[9,10] the particle-in-a-cell (PIC) method,[11-16] and path integral Monte Carlo (PIMC).[17,18] Each approach has its advantages and disadvantages: the Debye model is quick and easy but very approximate, quasi-harmonic lattice dynamics is very computationally intensive and neglects anharmonicity at constant volume, molecular dynamics is even more computationally intensive, usually requires small system sizes and neglects quantum occupation of phonon states, PIC requires supercells and neglects correlations between atomic motions and as has been implemented is classical, and PIMC with electrons can only be performed at very high temperatures and for small systems, although there may be tractable approaches that combine PIMC and density functional theory, perhaps using first-principles fitted potentials.[19] PIC is probably the fastest of the more accurate approaches for thermal properties, and seems particularly suited to properties at high (but not extreme) temperatures as occur in shock compression of metals. However, the accuracy of the PIC method has been recently called into question.[20] Here we test the accuracy by comparing with first-principles quasi-harmonic lattice dynamics while further investigating the properties of iron at high pressures and high temperatures.

Various thermodynamic and thermoelastic properties of iron at high pressure and temperature conditions have drawn significant experimental[21-35] and theoretical[11,13,36-43] attention. Iron is of great geophysical interest, because the Earth's core consists mainly of this element. Due to the extremely high temperatures (4000 to 8000 K) and high pressures (330 to 360 GPa) found in the Earth's inner core, it is difficult to probe these properties at such extreme conditions in an accurate way. Various first-principles theoretical calculations based on density functional theory and density function perturbation theory have been widely used to study material properties at these extreme conditions.

Early theoretical calculations only gave material properties at zero temperature, without any thermal effects included.[36, 41] One decade ago, Wassermann et al. first studied the thermal properties of iron at pressures up to 400 GPa and temperatures to 6000 K, using a first-principles fitted tight-binding total-energy model and the PIC approximation for the vibrational partition function.[11] As applied, the PIC model is a classical mean field approximation to the vibrational contributions to the free energy.[44, 45] The PIC model should be applicable to temperatures above the Debye temperature, but not too close to the melting temperature where the collective motions and diffusion become important. One crucial geophysical question is the origin of the elastic anisotropy found in the Earth's deep inner core, which might be associated with the lattice strain (especially the c/a axial ratio) of the $\varepsilon$-Fe phase. Wasserman et al. predicted a very rapid increase in the axial ratio with increasing temperature.[11] Stixrude et al. later applied the same method to examine the temperature and composition at the Earth's inner core conditions.[46] Instead of using the tight-binding model, Steinle-Neumann et al. applied the PIC model with a plane wave mixed basis pseudopotential method to examine the core's thermoelasticity and related aggregate properties.[13] All the three PIC studies agreed in predicting a strong increase in the axial ratio at high temperatures. Steinle-Neumann et al. attributed the seismological observations of the inner-core anisotropy to this increasing c/a axial ratio and the core's polycrystalline texture in which basal planes are partially aligned with the rotation axis.[13]

Alfe et al. presented density functional theory calculations to compute the free energies and other thermodynamics properties of $\varepsilon$-iron related to the Earth's solid inner core, and reached different conclusions by predicting a small increase in the c/a axial ratio with increasing temperature.[39, 40, 47] They predicted much smaller thermal effects on lattice strain than the earlier PIC studies did. They used a "frozen phonon" small displacement method to calculate the whole phonon spectrum in the harmonic approximation, and then added anharmonic corrections by either thermodynamic integration[39] or calculations of the thermal average stress from molecular dynamics simulation.[47] One advantage of these lattice dynamics calculations is that the results can be tested and validated by available experimental data on phonon dispersion and phonon densities of states.[30, 32] They also performed their own PIC calculations trying to test the quantitative accuracy of the PIC approximation and explain the big differences of the thermal effects on the lattice strain.[20] Their PIC results show some differences from both the earlier PIC and their frozen phonon calculations. They suggested that although PIC can be regarded as a way for calculating the geometric-mean harmonic frequency $\varpi$ and free energies, there is a constant factor difference between the $\varpi$ calculated from PIC and their frozen-phonon lattice dynamics. Gannarelli et al. performed first-principles molecular dynamics based on the density functional theory for hcp Fe at V= 47 bohr$^3$/atom,[47] and found a much smaller c/a axial ratio increase at high temperatures than the earlier PIC predictions.

Due to the importance of the issue and the large existing controversies, it is necessary to examine the thermal influences on the lattice strain using more accurate theoretical methods. With the exception of the projector-augmented-wave (PAW) method Gannarelli et al. used,[20, 47] most of the former calculations are based on pseudopotential, embedded atom or tight-binding total-energy models, where the validation and reliability of the results at extreme conditions may strongly depend on the development and versatility of the potentials or models. Here we examine the lattice dynamics and thermodynamics of ε-iron using the linear response full potential linear-muffin-tin-orbital (LMTO) method, where the results do not depend on the construction of the potentials and could provide benchmark to test the accuracy of the PIC model and other theoretical results. We also present our recent PIC calculations based on a tight-binding total-energy method, using a different integration method to make direct comparisons to the linear response results.

In Sec. II we detailed the theoretical techniques to obtain the thermal properties, as well as our linear response LMTO, PIC and tight-binding total-energy methods. In Sec. III we present the linear response and PIC results on the geometric mean vibrational frequencies, Helmholtz free energies and axial ratio of hcp Fe at high pressures and high temperatures, and make detailed comparisons to earlier PIC and other theoretical results. We conclude with a summary in Sec. IV.

## II. Theoretical methods

For many metals and alloys, the Helmholtz free energy F has three major contributions[48]

$$F(V,T)=E_{static}(V)+F_{el}(V,T)+F_{ph}(V,T) \qquad (1)$$

where V is the volume, and T is the temperature. $E_{static}(V)$ is the energy of a static lattice at zero temperature, $F_{el}(V,T)$ is the thermal free energy arising from electronic excitations, and $F_{ph}(V,T)$ is the phonon contribution. Both $E_{static}(V)$ and $F_{el}(V,T)$ can be obtained from first-principles calculations directly. We perform full potential LMTO calculations to evaluate $E_{static}(V)$ and $F_{el}(V,T)$ using multi-κ basis sets. Space is divided into the non-overlapping muffin-tin (MT) spheres surrounding each individual atom and the remaining interstitial region. The induced charge densities, the screened potentials and the envelope functions are represented by spherical harmonics within the MT spheres and by plane waves in the interstitial region. The self-consistent calculations are performed using 3κ-*spd*-LMTO basis set with one-center expansions performed inside the MT spheres.[49] The **k**-space integration needed for constructing the induced charge density is performed over a 12×12×12 grid. We use the Perdew-Burke-Ernzerhof (PBE) generalized-gradient approximation for the exchange and correlation energy.[50] When calculating $F_{el}(V,T)$, we assume temperature-independent eigenvalues for given lattice and nuclear positions, and only the occupation numbers change with temperature through the Fermi-Dirac distribution.[11]

We computed the phonons using the linear response method based on density functional perturbation theory and obtained the lattice vibrational contribution within the quasiharmonic approximation. We determined the dynamical matrix as a function of wave vector for a set of 28 irreducible **q** points at the 6×6×6 reciprocal lattice grid. The perturbative approach is employed for calculating the self-consistent change in the potential.[51, 52] Careful tests have been done against **k** and **q** point grids and many other parameters to make sure all the results are well converged. The phonon free energy $F_{ph}$ is obtained from the calculated phonon dispersion or phonon density of states[53]

$$F_{ph}(V,T) = k_B T \sum_{q,i} \ln[2\sinh\frac{\hbar\omega_i(q,V,T)}{2k_B T}] \tag{2}$$

In the PIC model, the partition function $Z_{cell}$ is approximated by calculating the energy of having one atom ("wanderer") move in the potential field of an otherwise ideal, fixed lattice[44, 45]

$$Z_{cell} = \lambda^{-3N} [\int_\Delta \exp[-\frac{U(\mathbf{r})-U_0}{k_B T}] d\mathbf{r}]^N \tag{3}$$

where $\lambda = h/(2\pi m k_B T)^{1/2}$ is the de Broglie wavelength of the atoms, $k_B$ is the Boltzmann constant, $U_0$ and $U(\mathbf{r})$ are the potential energies for the ideal system and for the system with the wanderer atom displaced by vector **r** from its equilibrium position, and N is the total number of atoms in the supercell. The integration is over the Wigner-Seitz cell $\Delta$, centered on the equilibrium position of the wanderer atom.

As described above, the vibrations are treated classically and PIC includes on-site anharmonicity. In order to compare the PIC and lattice dynamics results, it is useful to separate the harmonic and anharmonic contributions. Gannarelli et al. showed that this can be done by the series expansion of the perturbation energy in powers of ionic displacement, where the perturbation en-ergy is the potential energy difference between the distorted and the ideal lattice $\Delta U = U(\mathbf{r}) - U_0$. For the hcp symmetry, the perturbation energy can be expressed as[20]

$$\Delta U = \frac{1}{2!}[M\omega_a^2(r_x^2 + r_y^2) + M\omega_c^2 r_z^2] + \frac{1}{3!}K^{(3)}(r_y^3 - 3r_x^2 r_y) + \qquad (4)$$

$$\frac{1}{4!}[K_a^{(4)}(r_x^2 + r_y^2)^2 + K_{mix}^{(4)}(r_x^2 + r_y^2)r_z^2 + K_c^{(4)}r_z^4]$$

Here $r_x$ and $r_y$ are the Cartesian displacement components in the basal plane, and $r_z$ is the displacement along the hexagonal axis. The geometric-mean frequency $\varpi$ can be obtained from the harmonic vibrational frequencies in the basal plane $\omega_a$ and along the hexagonal axis $\omega_c$[20]

$$3\ln\varpi = 2\ln\omega_a + \ln\omega_c \qquad (5)$$

More theoretical details about the PIC model and its application to hcp Fe can be found in earlier publications.[11, 13, 20, 46]

## III. Results and Discussions

Ideally when comparing the accuracy between linear response and PIC, the calculations should be based on the same total-energy method. However, due to the large supercell size required in the PIC method, we use the tight-binding total-energy model, where the parameters for iron are determined by fitting to more than 4000 weighted input data consisting of the total energy and band structures of bcc, hcp and fcc iron over a large range of volumes, mainly from first-principles full potential linear-augmented-plane-wave (LAPW) calculations. The model has been widely used in many applications, where its reliability has been well testified.[11, 54-57] We further compared the calculated perturbation energies of moving the wanderer atom in a small 8-atom supercell, and the results for ε-Fe at volume of 60 bohr³/atom and c/a ratio of 1.6 are shown in Fig. 1. For all the three different directions where the wanderer is displaced, the results from the tight-binding and full potential LMTO methods are almost identical. Similar excellent agreements have been found for other volumes and axial ratios. Given the excellent agreement, it does not seem worth the much greater expense of performing self-consistent LMTO calculations for the larger cells.

It is important to perform careful convergence tests of the perturbation energy with respect to the supercell size and the number of **k** points used in the Brillouin zone integrations. In Fig. 2 we show the calculated $\Delta U$ for hcp iron at 60 bohr$^3$/atom and c/a ratio of 1.6, as the wanderer is displaced towards its nearest neighbor in the basal plane, for the 8-, 16-, 64- and 128-atom supercells respectively. The results for 64- and 128-atom supercells are almost identical, but in the 8- and 16-atom supercells, especially for the 8-atom cell, the results are clearly not converged. We obtained similar results for displacements in other directions and at other volumes and axial ratios. Thus we performed the PIC calculations on 64-atom supercell with periodic boundary conditions. We displaced the wanderer atom for several different directions, and calculated the perturbation energies at around ten different displacements along each direction. We then fitted the calculated 30~40 perturbation energies to Eq. (4), and obtained the geometric mean frequency and other fitted parameters. We numerically integrated Eq. (3) to get the vibrational free energy $F_{ph}$. We obtained the PIC results for ε-Fe at five different volumes from 40 to 80 bohr$^3$/atom, and varied the c/a axial ratio from 1.50 to 1.75 at each volume. The equilibrium axial ratio is determined by minimizing the Helmholtz free energies at a given temperature and volume.

As discussed in Sec. I, several earlier PIC studies of ε-Fe gave quite different conclusions for the c/a axial ratio changes at high temperatures. One major difference when applying the PIC model in these studies is how to do the 3-D integration over the Wigner-Seitz cell in Eq. (3). Wasserman et al. used the special direction integration method by taking advantage of the symmetry of the integrand. They expanded the integrand in orthogonal lattice harmonics, and then constructed the quadrature formula for the solid angle integration in such a way that it exactly integrates as many lattice harmonics as possible for the given number of directions.[11, 13, 46] Although they checked the convergence with respect to the number of special directions for the equation of state, the reliability of the simplified special direction integration method for strain energies is not very straightforward, especially considering that they only include three to four different special directions in their calculations. Gannarelli et al. simply fit their calculated first-principles perturbation energies to Eq. (4), and obtained the geometric mean frequencies and harmonic free energies from the fitted parameters, which can be used to compare with the lattice dynamics data directly.[20] In the current studies we employ similar numerical methods as Gannarelli et al. However, after fitting Eq. (4), we further calculated the PIC vibrational energy by integrating the Eq. (3) numerically.

The integrand in Eq. (3) decreases rapidly at large distances, as shown in Fig. 3, so the integration over the whole Wigner-Seitz cell can be further simplified. Wasserman et al. integrated over the inscribed sphere of radius equal to half of the nearest-neighbor separation,[11] and Gannarelli et al. chose the maximum displacements $\mathbf{r}_{max}$ in each direction so that the Boltzmann factor $\exp[-\Delta U(\mathbf{r}_{max})/k_B T] \approx 0.1$ at the maximum temperature of interest.[20] If given 6000 K as the maximum temperature, from Fig. 3 we can clearly see that the criterion used by Wasserman et al. is well justified , but the cutoff set up according to Gannarelli et al. is questionable. At both volumes of 70 and 50 bohr$^3$/atom, setting the cutoff at $\exp[-\Delta U(\mathbf{r}_{max})/k_B T] \approx 0.1$ is not sufficient to get well converged results at high temperatures. In our calculations, we used cutoff values in the same way as Wasserman et al. did, which guarantees our numerical integration over the whole Wigner-Seitz cell to be well converged.

The geometric-mean vibrational frequency $\varpi$ completely describes the harmonic free energy, so it is important to see how well the $\varpi$ calculated from the PIC model compares to the lattice dynamics data. Gannarelli et al. suggested that the PIC $\varpi$ differs from the $\varpi$ given by calculation of the full phonon spectrum by an almost constant factor over a wide range of volumes.[20] In contrast, our calculated $\varpi$ for hcp Fe at a given c/a ratio of 1.6 from PIC and linear response LMTO shows excellent agreement, as shown in Fig. 4, where we also include the computational data from Gannarelli et al.[20] Our computed frequencies from both PIC and linear response LMTO agree well with their "frozen phonon" lattice dynamics data, but not their PIC data. We do not know the exact reason, since too small cutoffs might change the anharmonic contributions significantly, but should have little influences on the harmonic properties. In Fig. 5 we compare the vibrational free energies calculated from quasiharmonic linear response LMTO and PIC methods. Again, we see excellent agreement between PIC and linear response lattice dynamics calculations. Since the PIC calculations include both the harmonic and anharmonic contributions, anharmonic effects prove to be small. At all the temperatures studied, anharmonic parts contribute less than 1.5% to the total phonon free energies.

In Figs. 4 and 5, it can be seen that there are some differences between PIC and linear response results for $\varepsilon$-Fe at 70 and 80 bohr$^3$/atom. At these large volumes, hcp is not the equilibrium structure for Fe. When applying the PIC model, the structure must be stable, otherwise both the special direction integration used by Wasserman et al.[11, 13] and the expansion of the perturbation energy as in Eq. (4) will fail since the undistorted structure will be a maximum or saddle point in the energy, rather than a minimum. We show the phonon frequencies at several selected **q** points calculated from our linear response LMTO method in Fig. 6. At large c/a ratios, several phonon branches show strong softening, which makes the hcp structure dynamically unstable. At V= 60 bohr³/atom, hcp Fe becomes unstable before approaching c/a ratio of 1.8. The PIC model might show large errors as the structure approaches lattice instability.

On-site anharmonicity is included in the PIC model, and its free energy contributions can be expressed as[20]

$$f_{anharm}^{PIC} = dT^2 + O(T^3) \qquad (6)$$

Gannarelli et al. gave an expression for the anharmonic coefficient $d$ which they claimed was exact.[20] However, their equation gives the wrong unit for $d$. We obtained $d$ by first calculating the anharmonic free energies through numerical integration and then fitting the energies to Eq. (6). For all the different volumes and axial ratios, the anharmonic free energies can be well fitted as a parabola in temperature. We show the calculated anharmonic coefficient at several different volumes in Fig. 7, in comparison to a couple of earlier theoretical predictions. Gannarelli et al. obtained positive anharmonic free energy through PIC,[20] while their vibrationally correlated calculations gave negative values.[39] Gannarelli et al. attributed the discrepancy to the PIC approximation, and concluded that PIC gave a completely incorrect account of anharmonicity. However, our current PIC anharmonic coefficients show good agreement with the vibrationally correlated calculations, in both the sign and the values. The small cutoff, instead of the PIC model itself, might account for the different anharmonic behaviors observed in the early PIC study. It should be noted that the anharmonic contributions are very small compared to the harmonic energies, and the anharmonic effects on the thermal equation of state and c/a strain are almost negligible for iron under the conditions of interest here.

In Fig. 8 we show the calculated equilibrium axial ratio of hcp Fe at room temperature as a function of pressure. Since the temperature is low here, the thermal effects are small, and the results from the PIC model and linear response lattice dynamics are almost identical. Both calcula-tions predict a slight increase of the c/a ratio with increasing pressure, and are in excellent agreements with earlier theoretical predictions[47] and several diamond-anvil-cell experimental data.[34, 58, 59] The axial ratio calculated from both PIC and linear response at volume of 50 bohr$^3$/atom show a small increase with increasing temperature, although PIC calculations trend to give a slightly larger c/a ratio increase, as shown in Fig. 9. The calculated axial ratios at high temperatures agree well with recent theoretical studies[47] and in situ synchrotron X-ray diffraction experiments.[34] The present results are significantly different from some earlier PIC calculations, where the axial ratio was predicted to increase very rapidly with temperatures at the similar volumes.[13] Our current study suggests that this is not due to the error in the PIC model itself, which could describe the classical vibrational behavior accurately. The errors in the earlier PIC studies must come from the application of the special directions method for 3D integration, or the PIC errors due to the lattice instability at large c/a ratios. Since one earlier PIC study focusing on the Earth's core conditions gave large axial ratio increase,[13] where ε-Fe proves to be quite stable, the PIC errors introduced by lattice instability are not the major reason.

## IV. Conclusions

In summary, we have performed detailed first-principles linear response lattice dynamic and particle-in-cell model calculations to study the properties of ε-Fe at high pressures and high temperatures, and to test the accuracy of the PIC model. The tight-binding model used in the PIC study are fitted to the full potential LAPW data, and the calculated perturbation energies to move the wanderer atom in the supercell show excellent agreements with the full potential LMTO calculations. The PIC model gives good agreement with linear response LMTO results for the stable hcp structure, since the calculated geometric mean frequency, the Helmholtz free energy and the axial ratio as a function of temperature and pressure all agree well. PIC calculations include on-site anhar-monicity, which proves to be small for ε-Fe. Overall, the PIC model describes the classical vibrational behavior of iron quite accurately, but it might fail when the structure approaches lattice instability; ε-Fe becomes dynamically unstable at large c/a ratio at low pressures. The large controversies about several earlier PIC studies are not introduced by the PIC method itself, but must come from errors in its application.

**ACKNOWLEDGEMENTS**

We thank G. Steinle-Neumann for useful discussions and thank S. Savrasov for use of his full-potential LMTO code. This work was supported by DOE ASC/ASAP subcontract B341492 to Caltech DOE w-7405-ENG-48. Computations were performed on the Opteron Cluster at the Geophysical Laboratory and the ALC cluster at the Lawrence Livermore National Laboratory, supported by DOE and the Carnegie Institution of Washington.

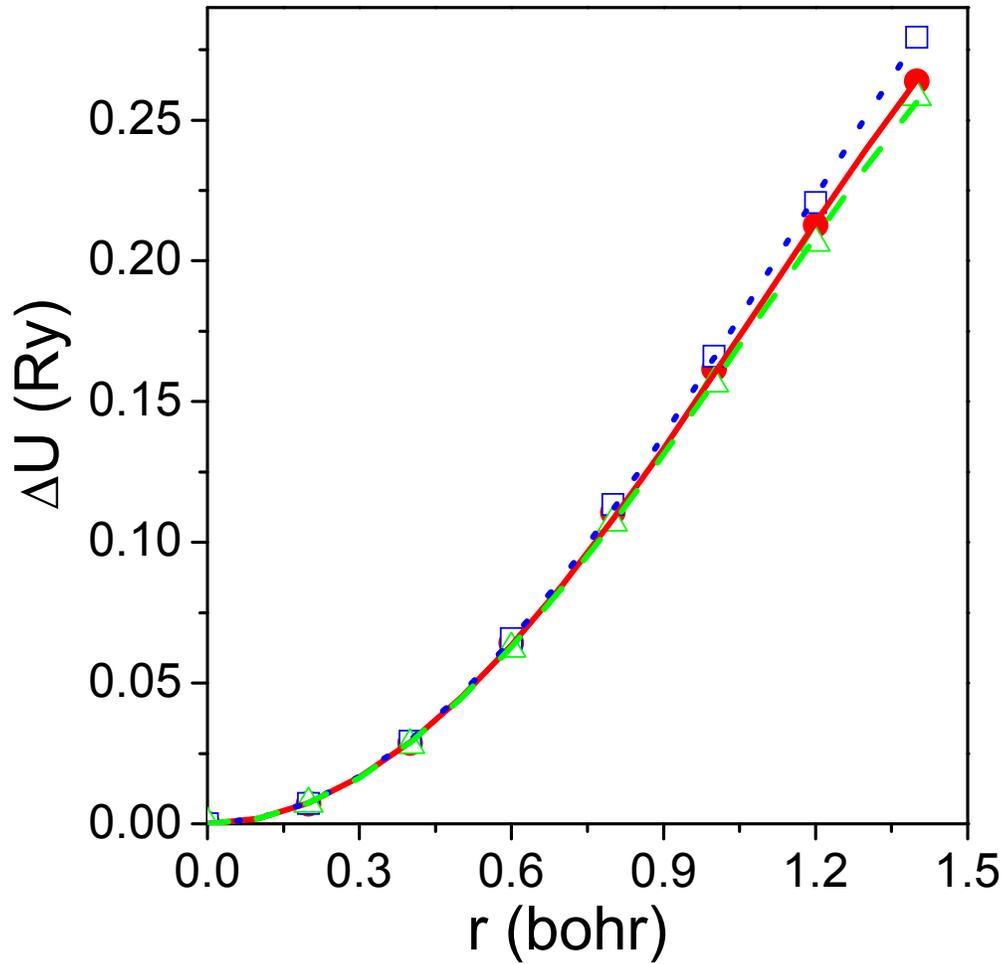

Fig. 1. The calculated perturbation energy ΔU for displacing the wanderer atom in a small 8-atom supercell of ε-Fe at 60 bohr$^3$/atom and c/a ratio of 1.6. The results are shown for the wanderer to move in the [100] direction in the basal plane (solid line for tight binding and filled circles for LMTO), [010] direction in the basal plane (dashed line for tight binding and open triangles for LMTO), and [001] direction along the hexagonal axis (dotted line for tight-binding and open squares for LMTO). The tight-binding and LMTO results show excellent agreement for all the directions.

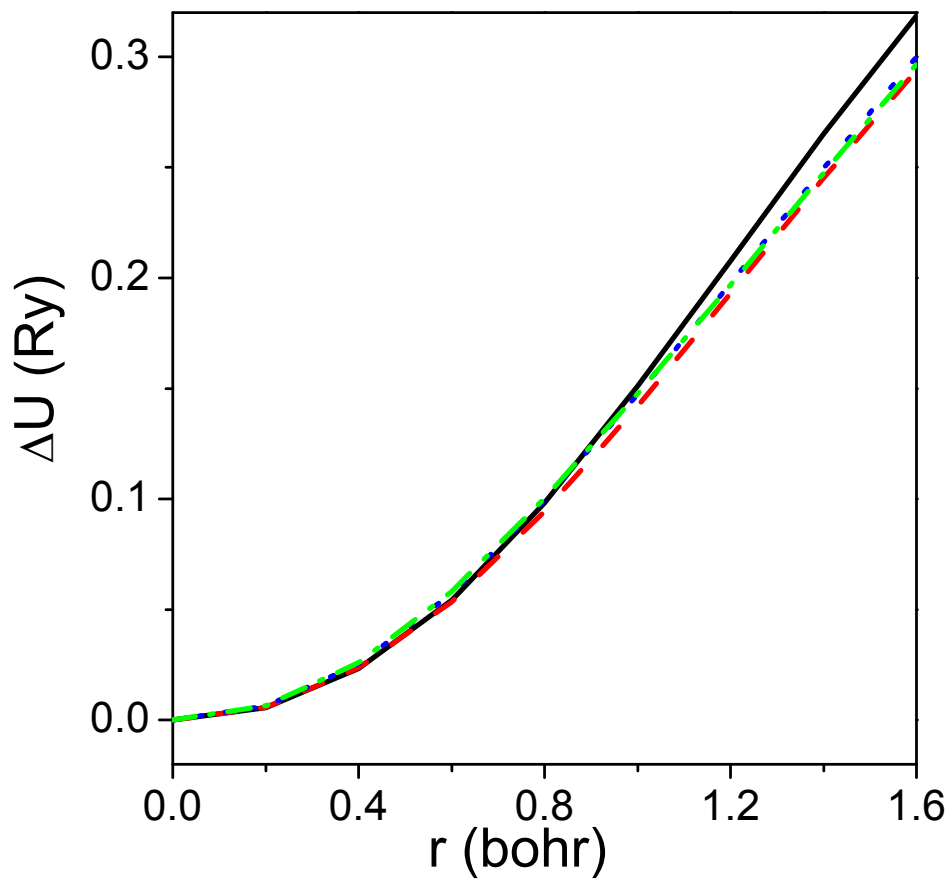

Fig. 2. Dependence of the perturbation energy ΔU on the supercell size for hcp iron at 60 bohr$^3$/atom and 1.6 c/a ratio, where the wanderer atom is displaced towards its nearest neighbor in the basal plane. The supercell contains 8 (solid line), 16 (dashed line), 64 (dotted line) and 128 (dotted dash line) atoms, respectively. The results for 64 and 128 atoms supercell are almost identical.

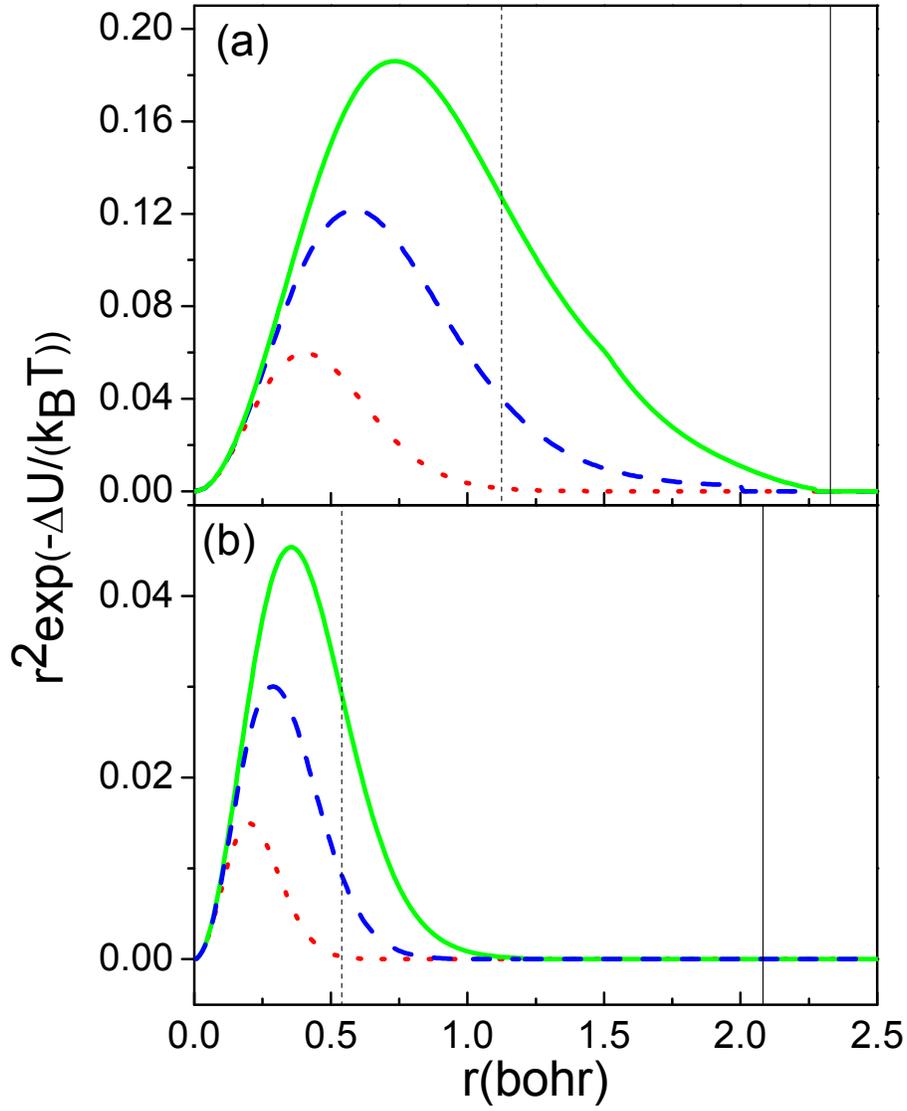

Fig. 3. Integrand in Eq. (3) along the nearest neighbor direction in the basal plane for hcp Fe at c/a ratio of 1.6 and volumes of (a) 70 and (b) 50 bohr$^3$/atom, and at temperatures T= 2000 K (dotted line), 4000 K (dashed line) and 6000 K (solid line). The integrand decays rapidly with increasing radius. Solid vertical lines show the cutoff according to Wasserman et al. (Ref. 11), and the dotted vertical line is the cutoff setting from Gannarelli et al. (Ref. 20)

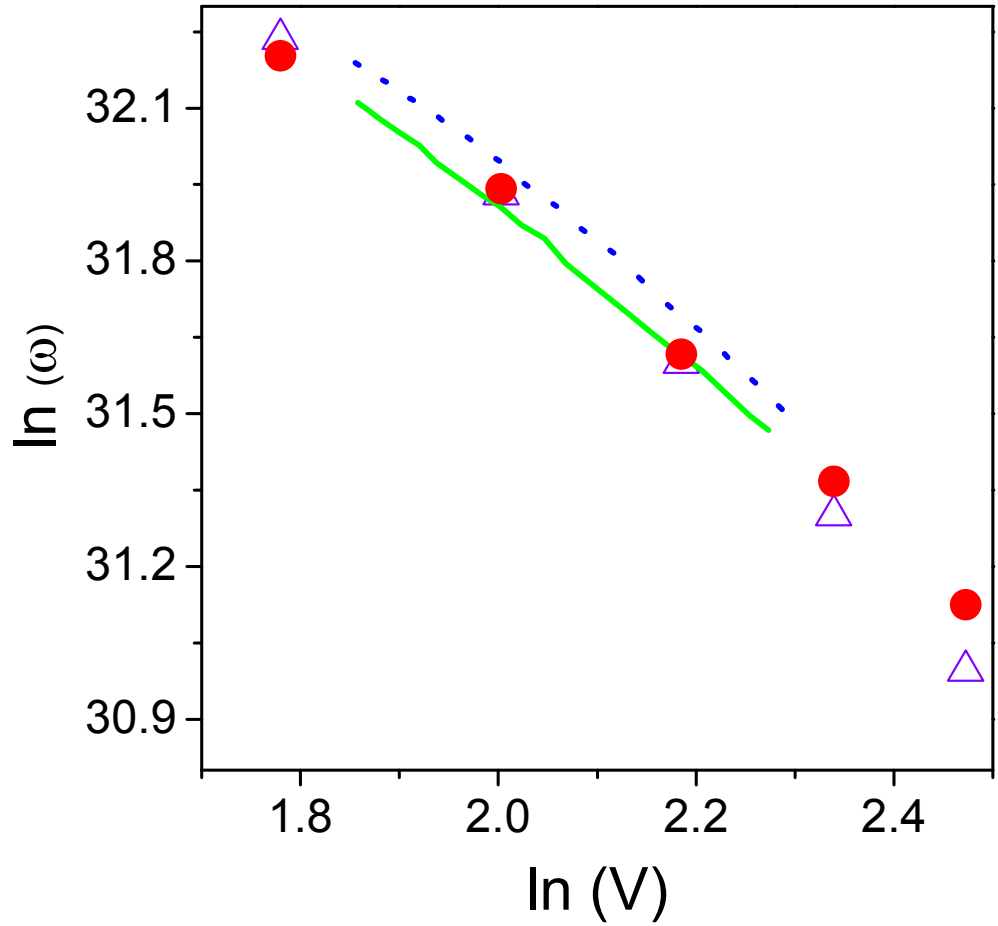

Fig. 4. The calculated geometric-mean vibrational frequencies from PIC (open triangles) and linear response (filled circles) agree well for hcp Fe under pressures, different from earlier theoretical predictions (lattice dynamics and PIC results are shown as solid and dotted lines, Refs. 40 and 20). The differences between PIC and linear response results increase as the structure approaches lattice instability.

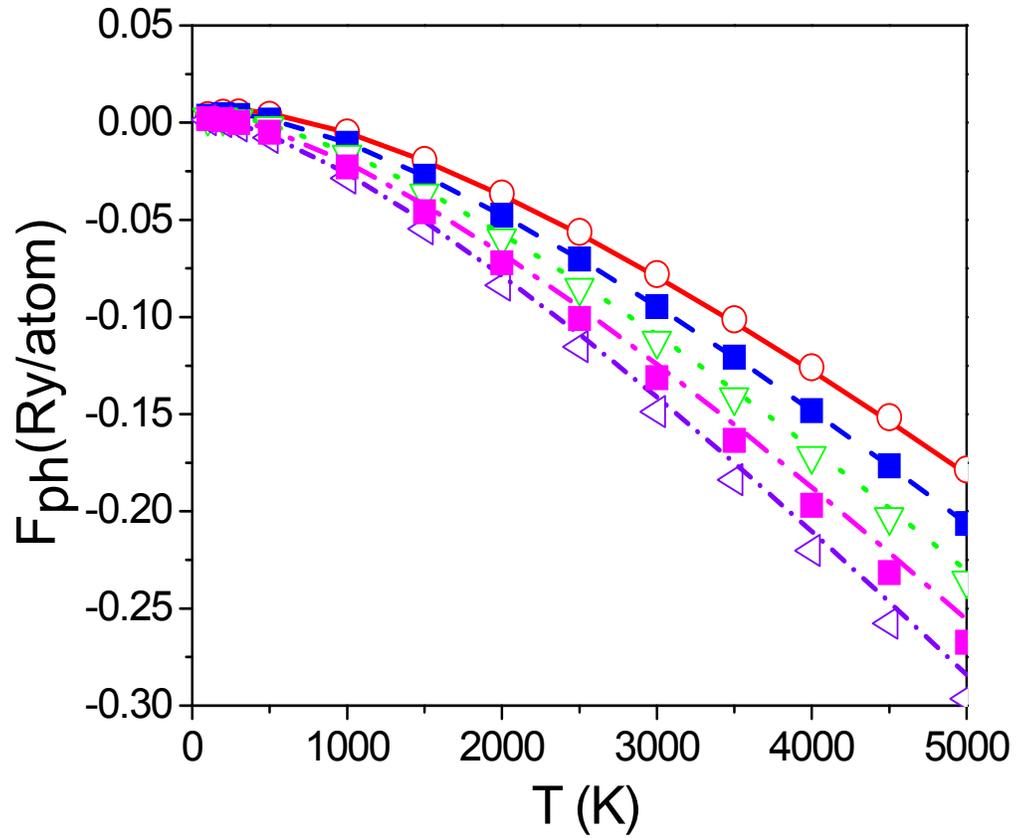

Fig. 5. The calculated vibrational free energies $F_{ph}$ as a function of temperature at volumes from 40 (top curve) to 80 (bottom curve) bohr$^3$/atom with 10 bohr$^3$/atom interval. The linear response (curves) and the PIC (symbols) results agree well for hcp Fe under pressures. The differences increase when the structure approaches instability.

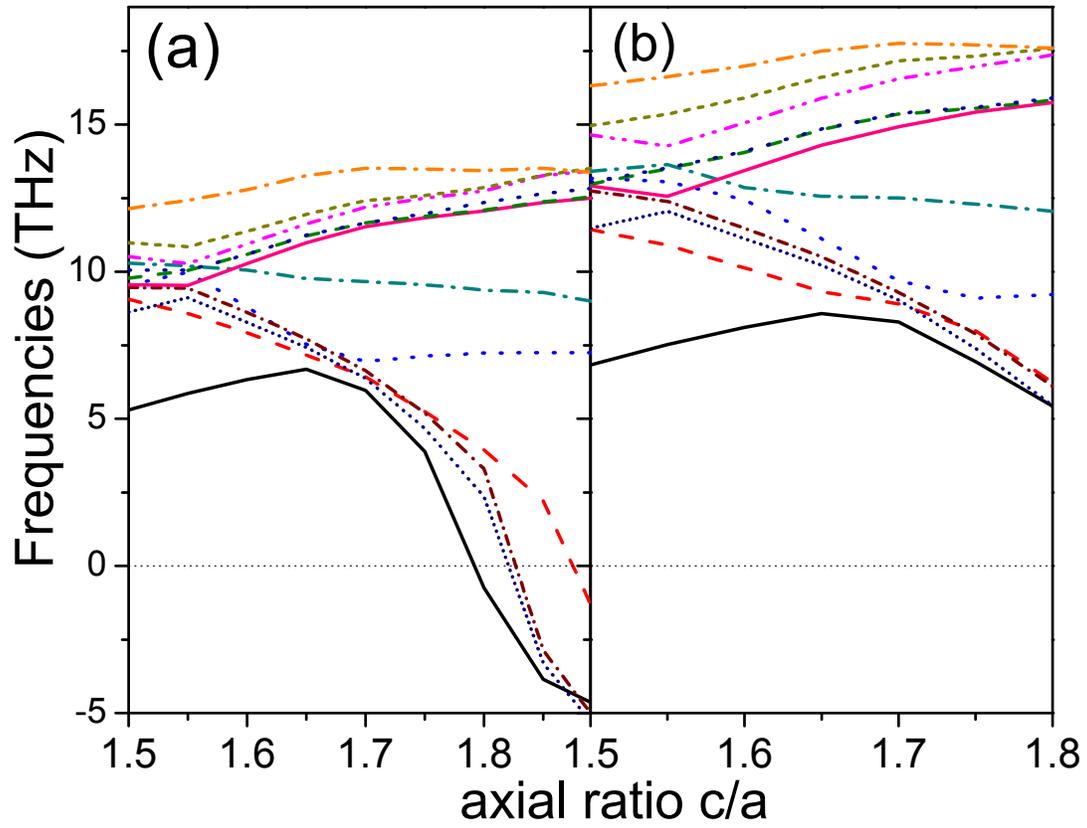

Fig. 6. The calculated phonon frequencies at several selected **q** points, at a function of c/a ratio at volumes of (a) 50 and (b) 60 bohr$^3$/atom from linear response LMTO study. The results show the lattice instability of ε-Fe at large c/a ratios, especially at low pressures.

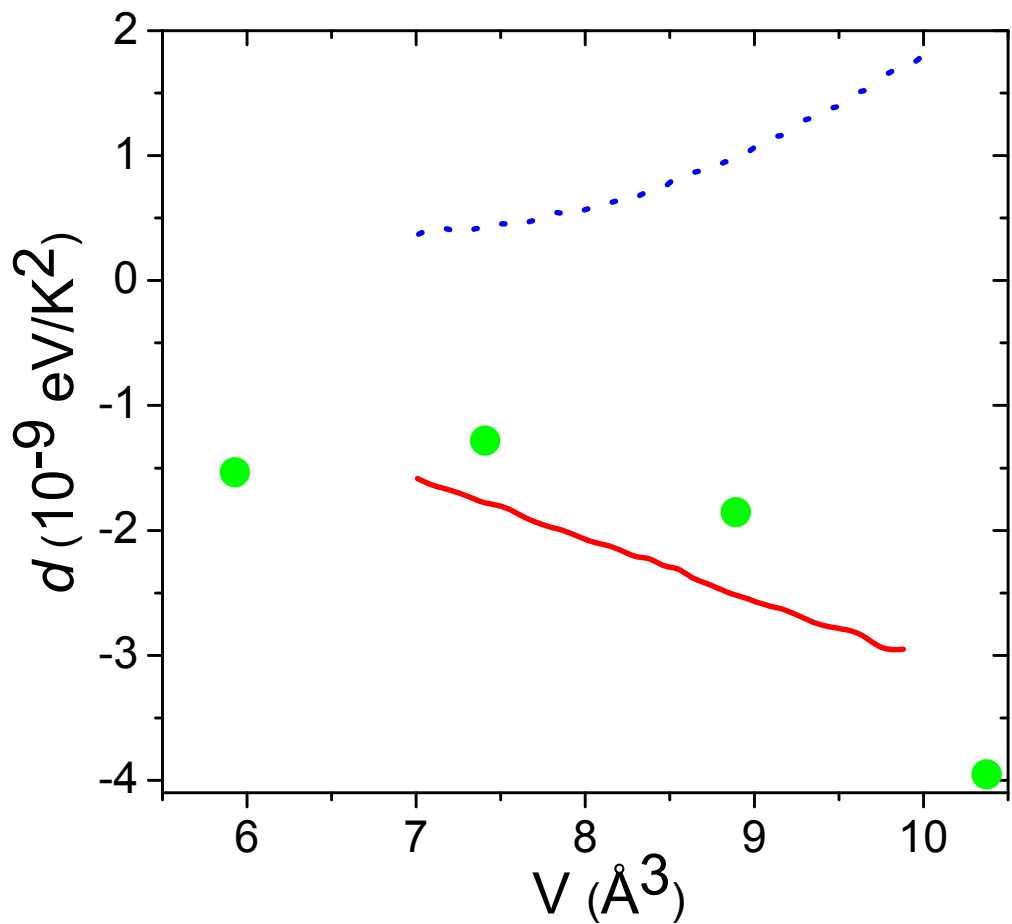

Fig. 7. Anharmonic coefficient *d* as a function of atomic volume. The current PIC results are given by filled circles, and the solid and dotted lines represent the results form the vibrational correlated (Ref. 39) and earlier PIC (Ref. 20) calculations.

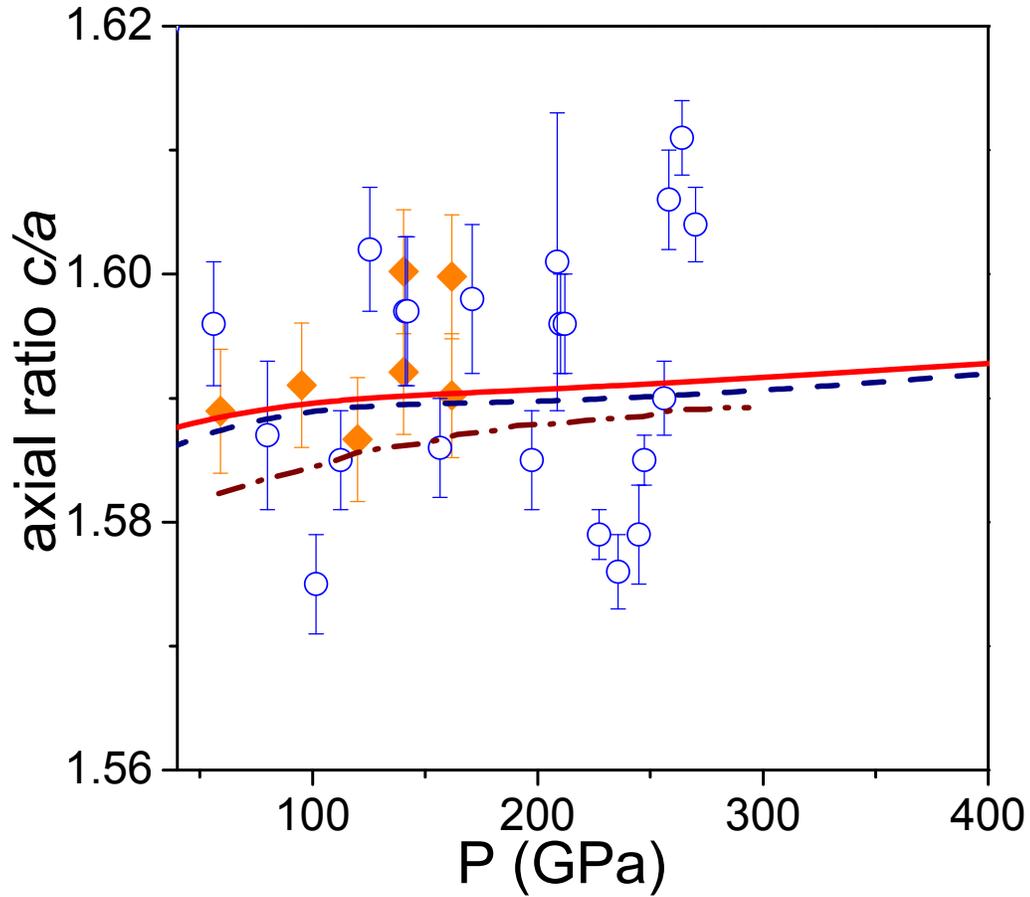

Fig. 8. Calculated equilibrium axial ratio of hcp Fe as a function of pressure at ambient temperature. Solid and dashed curves show the results calculated from current PIC and linear response LMTO calculations, respectively. These are in good agreement with earlier theoretical PAW results (dotted dash line, Ref. 47), as well as the experimental data (filled diamonds with error bars, Ref. 34; open circles with error bars, Refs. 58 and 59), usually within the quoted error bars.

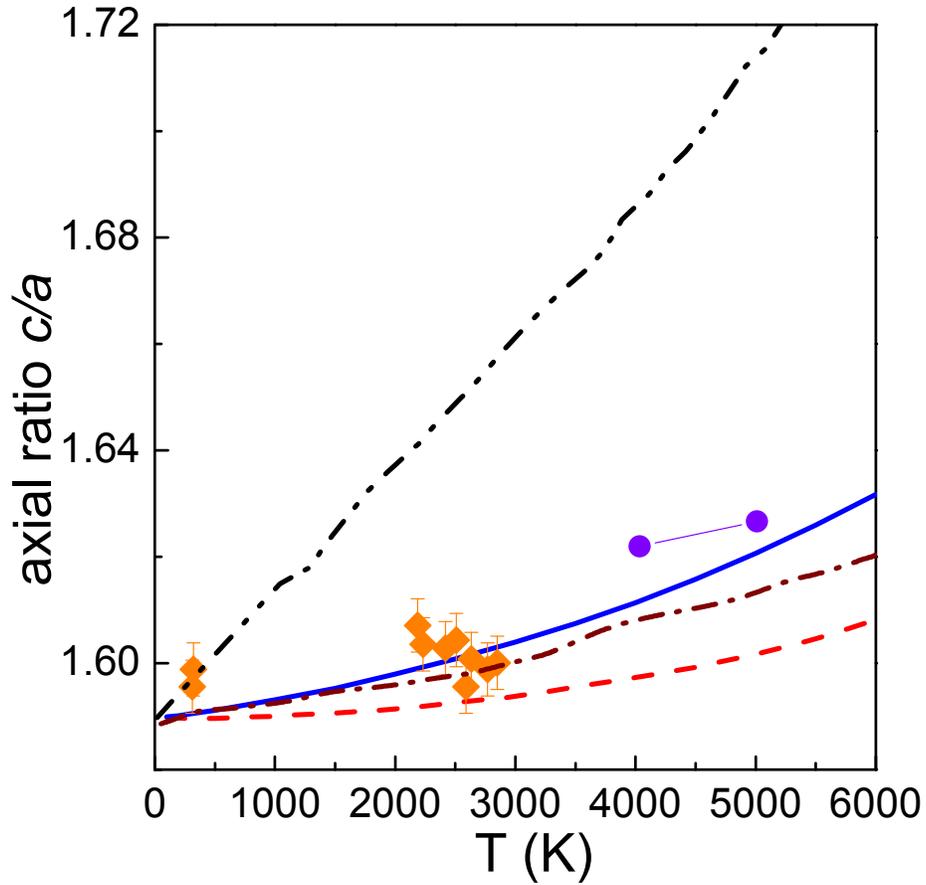

Fig. 9. The calculated axial ratios of ε-Fe as a function of temperature. The solid and dashed lines are current PIC and linear response results for hcp Fe at 50 bohr$^3$/atom, respectively. Also shown are some recent first-principles data using molecular dynamics (filled circles, V=47 bohr$^3$/atom, Ref. 47), lattice dynamics (dot dashed line, V= 47 bohr$^3$/atom, Ref. 47) and PIC model (dot dot dashed line, V= 50 bohr$^3$/atom, Ref. 13), as well as the in situ X-ray diffraction experimental data (squares with error bars, V= 52 bohr$^3$/atom, Ref. 34).